\documentclass[conference]{IEEEtran}
\IEEEoverridecommandlockouts
\usepackage{cite}
\usepackage{amsmath,amssymb,amsfonts}
\usepackage{algorithmic}
\usepackage{graphicx}
\usepackage{float}
\usepackage{subfigure}
\usepackage{textcomp}
\usepackage{xcolor}
\usepackage{multirow}
\def\BibTeX{{\rm B\kern-.05em{\sc i\kern-.025em b}\kern-.08em
    T\kern-.1667em\lower.7ex\hbox{E}\kern-.125emX}}
\begin{document}
	
	\title{Efficient Accelerator for Dilated and Transposed Convolution with Decomposition}
	
\author{\IEEEauthorblockN{Kuo-Wei, Chang, and Tian-Sheuan Chang}
	\IEEEauthorblockA{\textit{Dept. of Electronics Engineering, 
			National Chiao Tung University
			Hsinchu, Taiwan}
		}
	\thanks{K. -W. Chang and T. -S. Chang, "Efficient Accelerator for Dilated and Transposed Convolution with Decomposition," 2020 IEEE International Symposium on Circuits and Systems (ISCAS), 2020, pp. 1-5, doi: 10.1109/ISCAS45731.2020.9180402.}	}
	
	\maketitle
	
	\begin{abstract}
Hardware acceleration for dilated and transposed convolution enables real time execution of related tasks like segmentation, but current designs are specific for these convolutional types or suffer from complex control for reconfigurable designs. This paper presents a design that decomposes input or weight for dilated and transposed convolutions respectively to skip redundant computations and thus executes efficiently on existing dense CNN hardware as well. The proposed architecture can cut down 87.8\% of the cycle counts to achieve 8.2X speedup over a naive execution for the ENet case.	
	\end{abstract}
	
	
	\section{Introduction}
	Convolution neural networks (CNNs) based image segmentation \cite{RCNN, enet} has been widely used in scene understanding, medical purposes, and action recognition during recent years for its significant improvement over traditional approaches. However, the computation of CNNs requires billions of multiplications and accumulations (MACs). Thus, hardware acceleration for CNNs is demanded to provide high parallelism for high throughput to achieve real time execution.
	
\begin{figure}[t]
	\centering{\includegraphics[height=60mm]{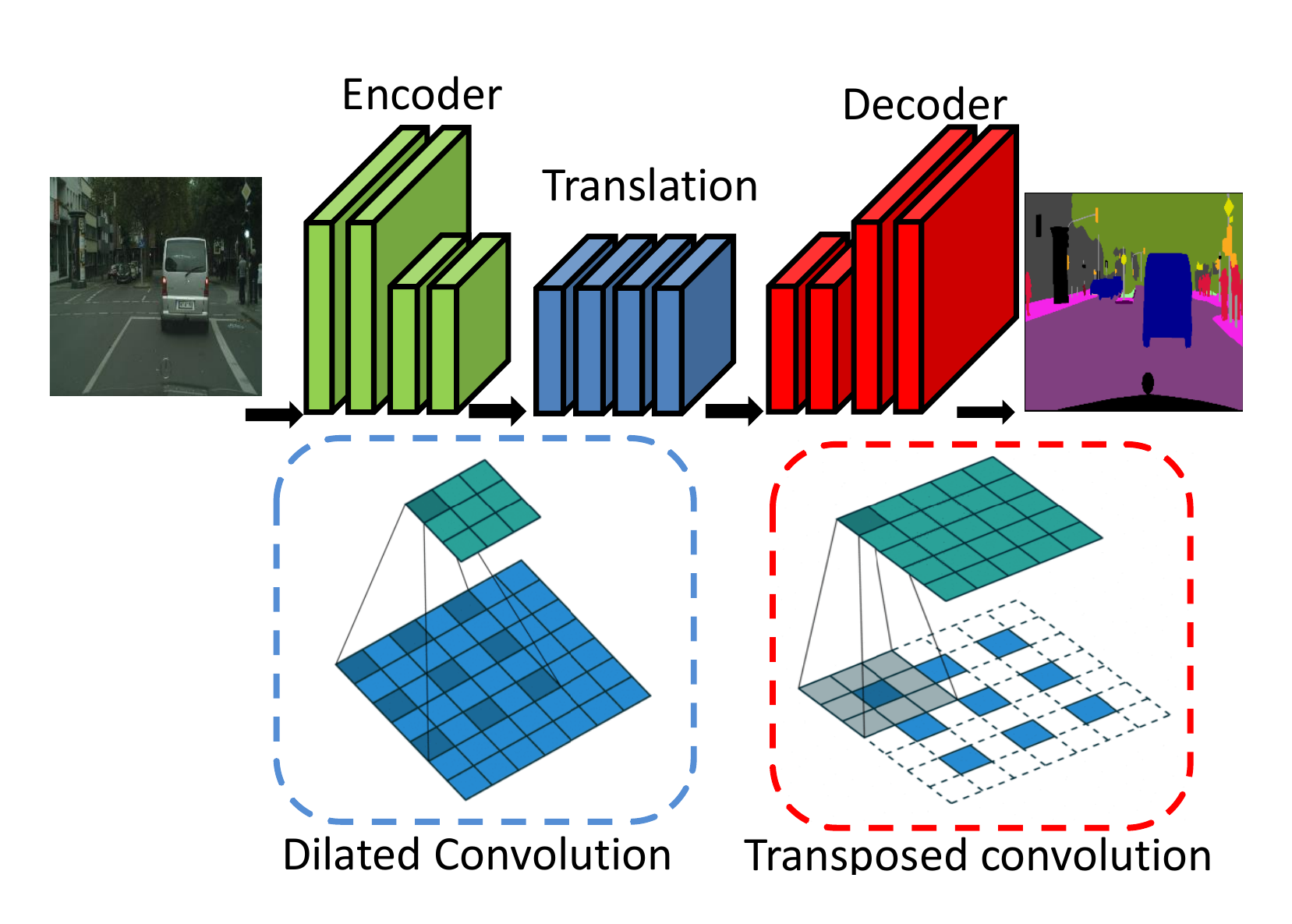}}
	\caption{The architecture of image segmentation.}
	\label{fig:seg_example}
\end{figure}

	Various hardware accelerators \cite{eyeriss,dna,DTCNN,USCA,GAN_ACC,DNPU} have been proposed recently. Typical accelerators focus on widely used convolutions with stride one (e.g. 3$\times$3), called dense CNN in this paper\cite{eyeriss,dna}. \cite{eyeriss} adopts a spatial array architecture and row stationary data flow for classification \cite{alexnet,vgg,resnet,googlenet_1,googlenet_2,mobilenet}. \cite{dna} proposes a systolic array architecture with full reconfigurations for different convolutional kernels on classification. \cite{DNPU} uses different hardware units to support CNNs and recurrent neural networks. For acceleration of segmentation, a typical segmentation consists of dilated and transposed convolutions that have many zeros at the weight or input as shown in Fig.~\ref{fig:seg_example}, which results in sprase CNN and low hardware utilization when naïvely mapped to a typical dense CNN hardware accelerators. Thus, for accelerators of dilated and transposed convolutions, \cite{DTCNN} proposes an accelerator with delay cells to support dilated and transposed convolution in the segmentation. \cite{USCA} provides a unified systolic array to accelerate different types of convolution. \cite{GAN_ACC} uses a cascading filter structure to support transposed convolutions for generative neural networks \cite{dcgan}. However, most of existing accelerators are tailored for a specific task, which will incur high reconfiguration hardware cost to support different tasks.
	
	To support dilated and transposed convolutions without high reconfiguration costs, this paper proposes to decompose input or weight for dilated and transposed convolutions respectively such that all these convolutions are reduced to normal dense CNN and easily executed on existing dense CNN hardware with no overhead. Applying this flow to a dense CNN design \cite{VWA} does not need extra logic and can save 97\% and 71\% of cycle count for dilated and transposed convolutions in ENet\cite{enet}, respectively. 
	
	\section{Proposed Method}
	\subsection{Overview of Segmentation}
	 Fig.~\ref{fig:seg_example} shows the network architecture of segmentation that consists of an encoder, translation, and decoder. The encoder is a typical CNN with convolutions layers and pooling layers to extract high level features. Then these features are further processed with dilated convolutions that use the enlarged kernels with zero insertion to keep feature map size unchanged in the translation part. These feature maps are then upsampled to generate output with the same size as input with the transposed convolutions in the decoder. The  transposed  convolution  enlarges  input by inserting zeros between the adjacent input elements and convolves with a normal kernel to generate enlarged output. Both dilated and transposed convolutions contain a large number of zero computations. How to skip these zero computations without complex control cost is a challenging task for hardware accelerators.

	\begin{figure}[t]
		\centering{\includegraphics[height=30mm]{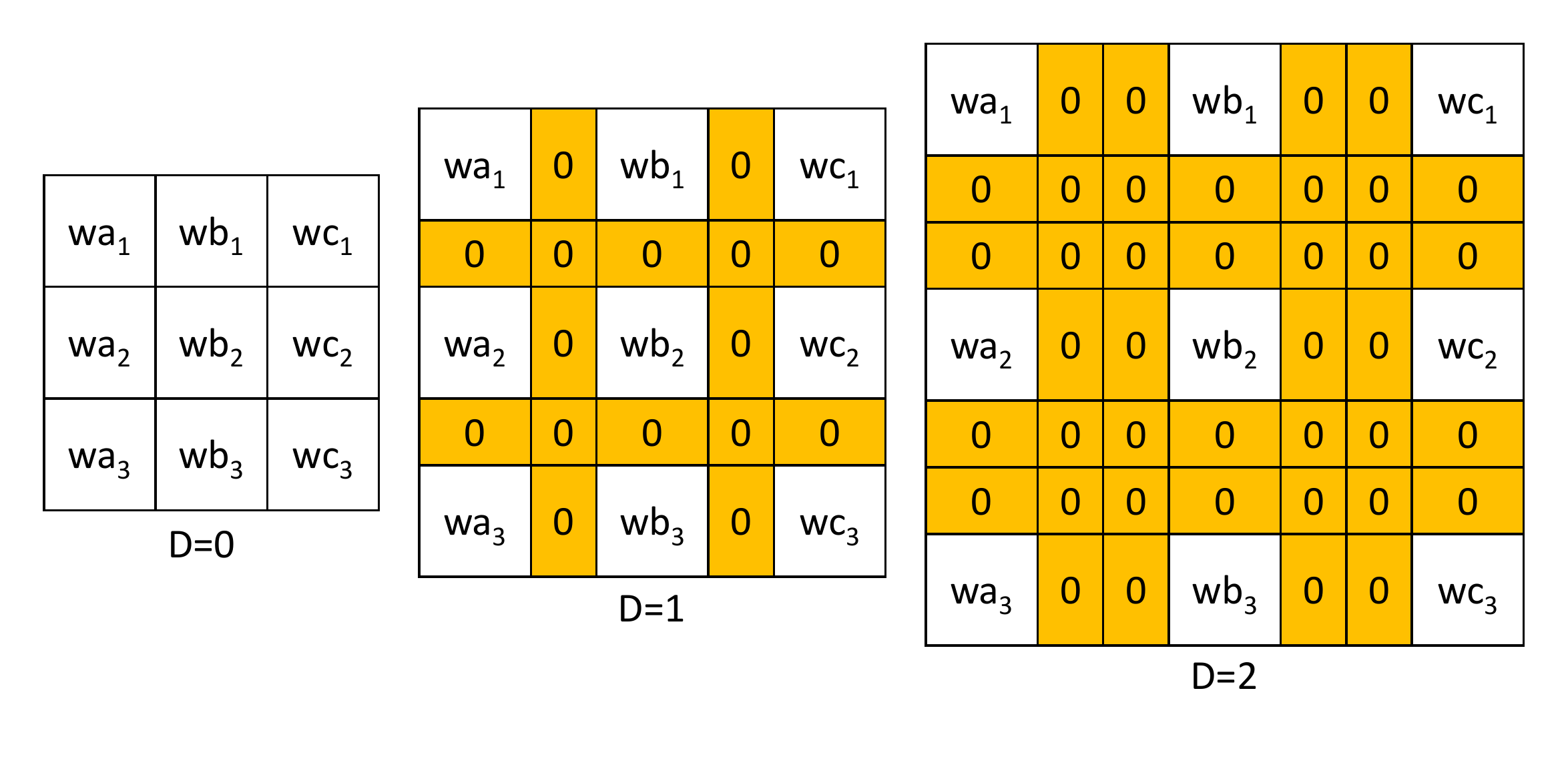}}
		\caption{Weight matrices with different $D$ in dilated convolutions. The size equal to $(2 \times D + 3) \times (2 \times D + 3)$.}
		\label{fig:dilated_weight}
	\end{figure}
	\begin{figure}[t]
		\centering{\includegraphics[height=35mm]{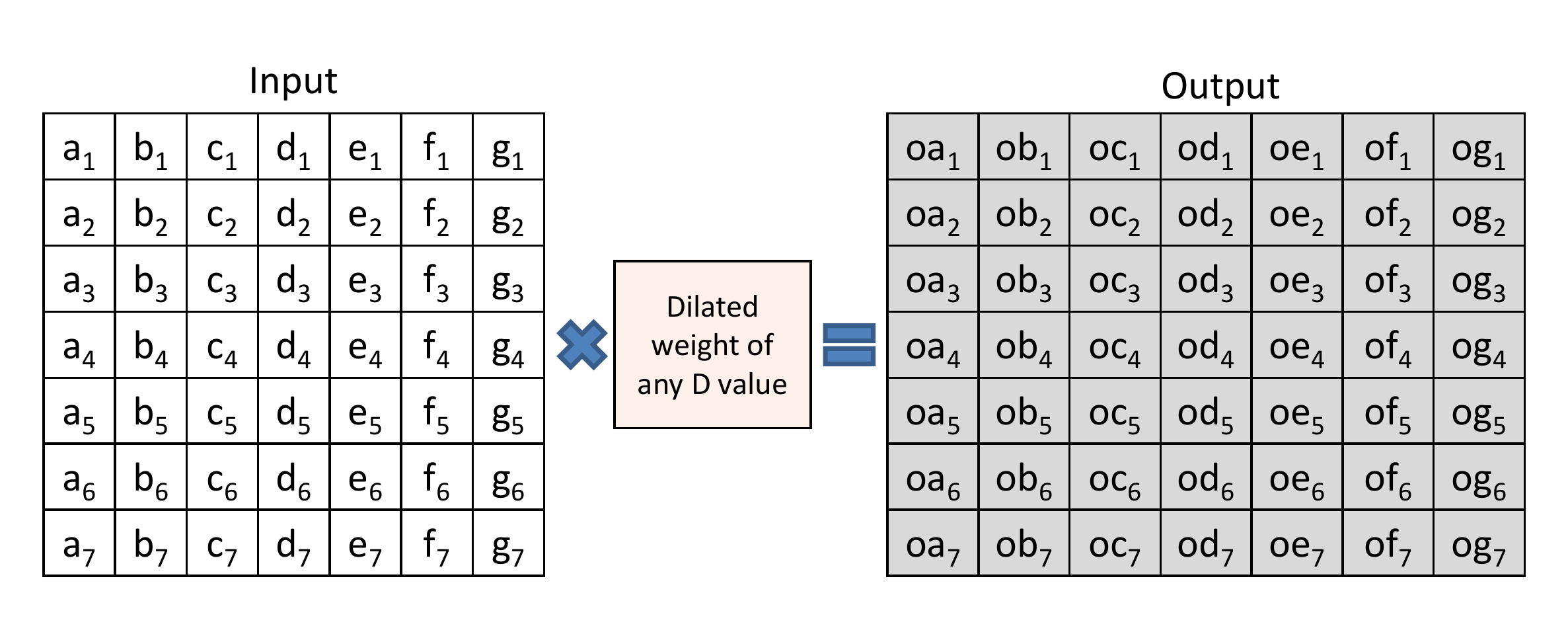}}
		\caption{A dilated convolution example with 7$\times$7 input and dilated weight of any $D$ to generate 7$\times$7 output. In which, $1+D$ zeros are padded around input to maintain output size the same as input size.}
		\label{fig:example_dilated}
	\end{figure}
	\begin{figure}[t]
		\centering{\includegraphics[height=40mm]{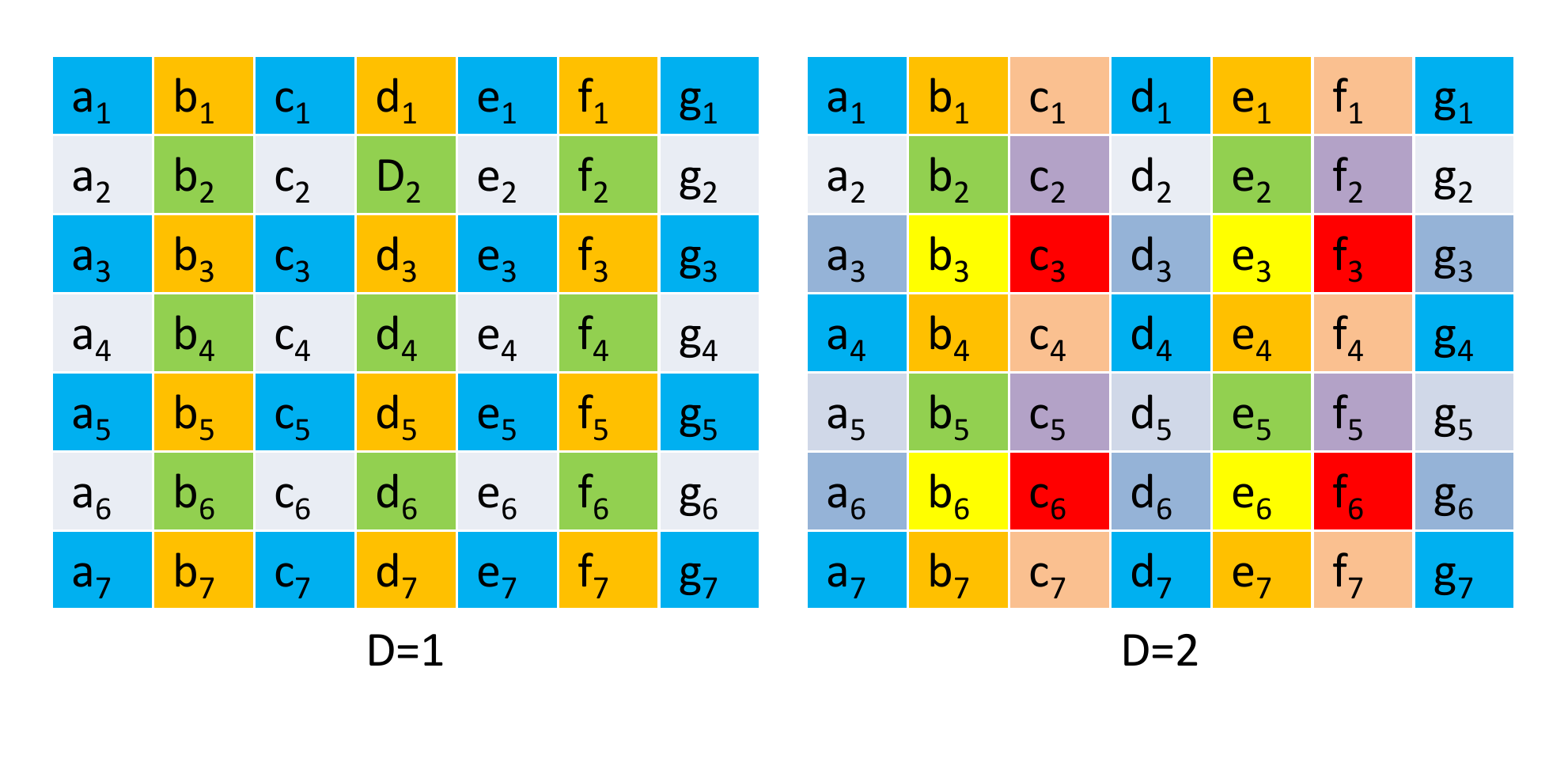}}
		\caption{The decomposition for dilated convolutions. Input are decomposed to 4 and 9 blocks (denoted with different colors) at $D=1$ and $D=2$, respectively. The input elements with the same color are in the same block.}
		\label{fig:decom_dilated}
	\end{figure}
	
	\begin{figure}[t]
		\centering{\includegraphics[height=40mm]{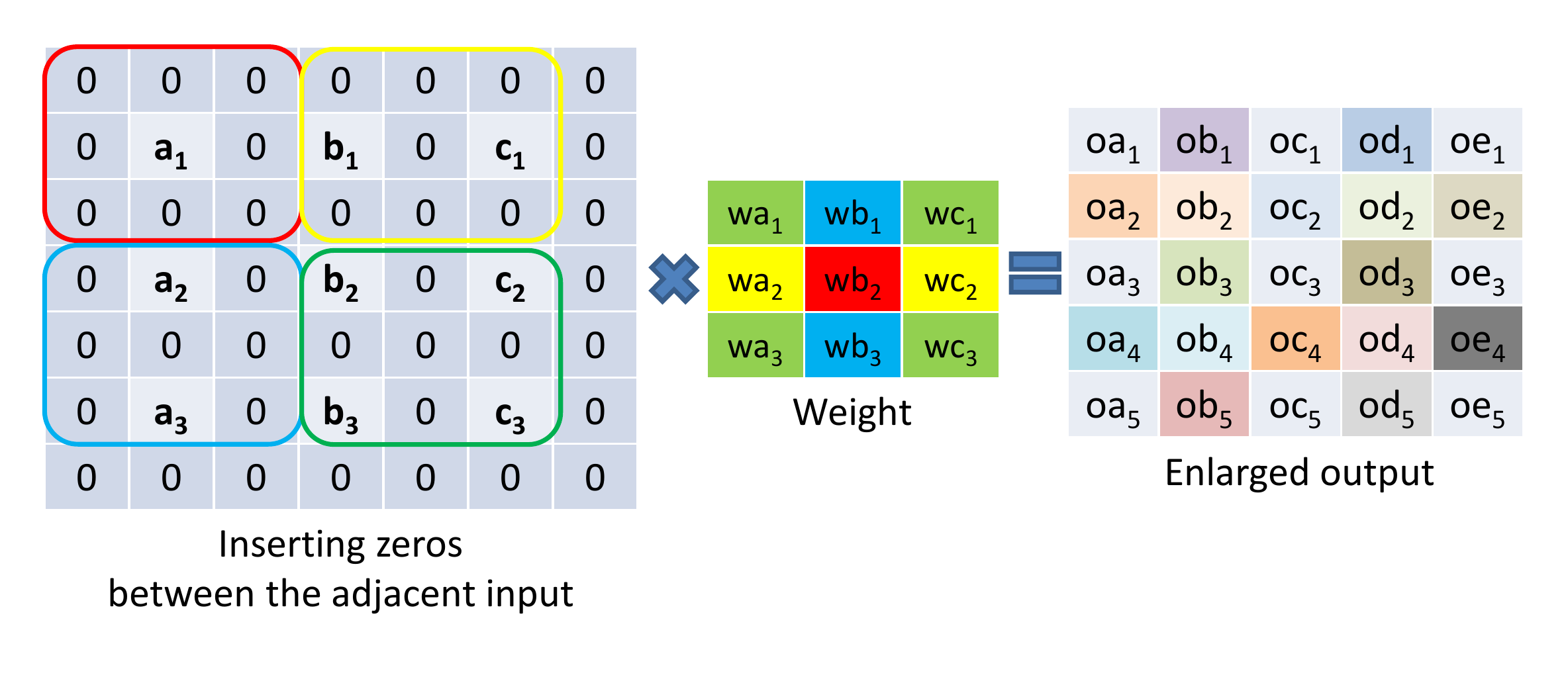}}
		\caption{A transposed convolution example with 3$\times$3 input and 3$\times$3 to generate enlarged 5$\times$5 output by inserting zeros between the adjacent input. Only four kinds of nonzero computations exists as denoted with different color squares.}
		\label{fig:deconv}
	\end{figure}
	\begin{figure}[t]
		\centering{\includegraphics[height=25mm]{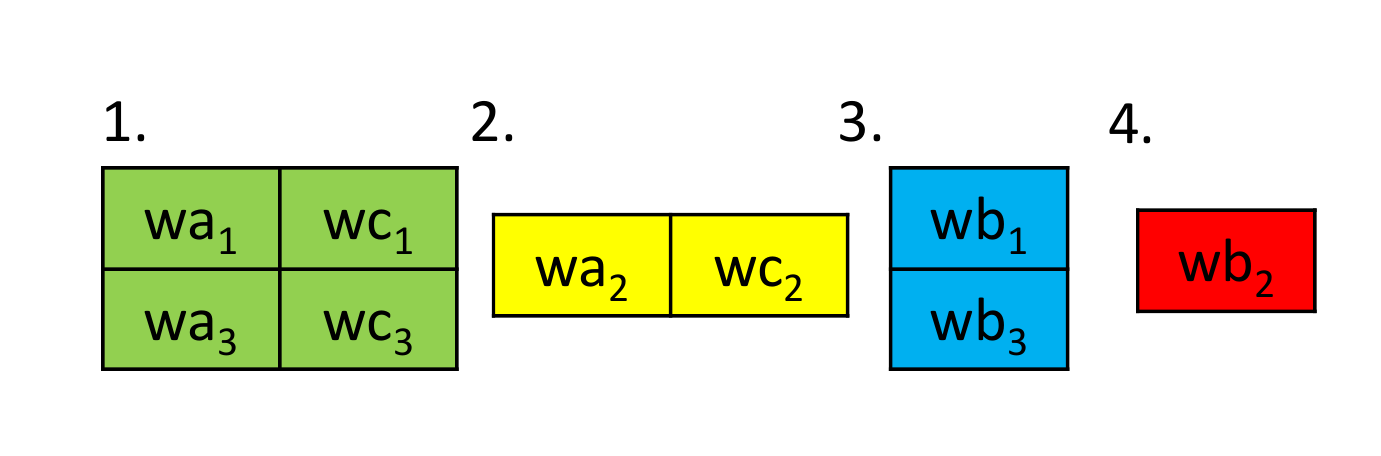}}
		\caption{The decomposition for transposed convolutions. 3$\times$3 weight matrices are decomposed to four blocks.}
		\label{fig:deconv_decom}
	\end{figure}
	\begin{figure}[t]
		\centering{\includegraphics[height=80mm]{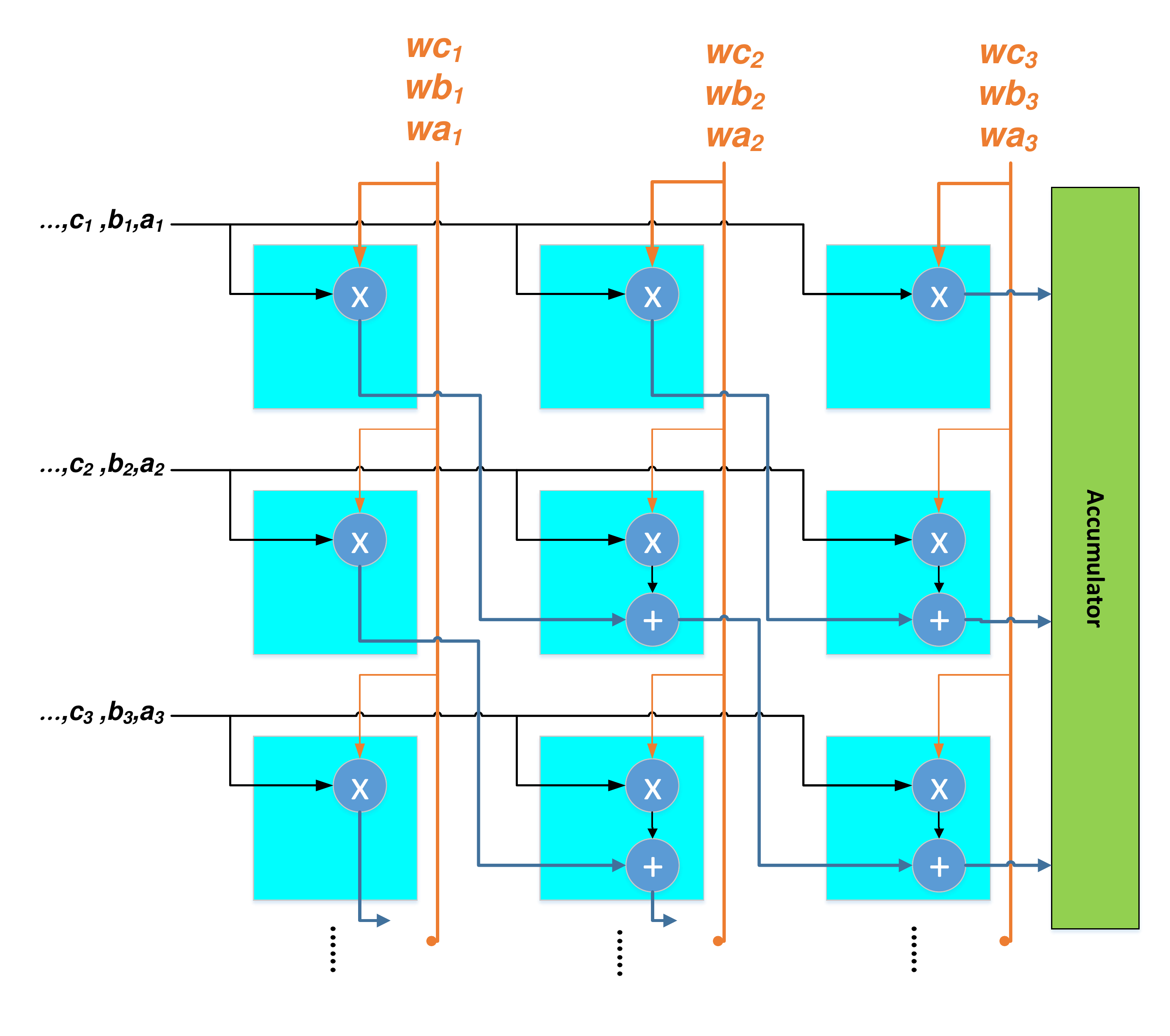}}
		\caption{The architecture of the PE block. In which, input column vectors and weight column vectors
		 are broadcasted to a $n\times3$ MAC array to generate partial sum or output data to accumulator. \cite{VWA}}
		\label{fig:PE}
	\end{figure}
	\begin{figure}[t]
		\centering{\includegraphics[height=48mm]{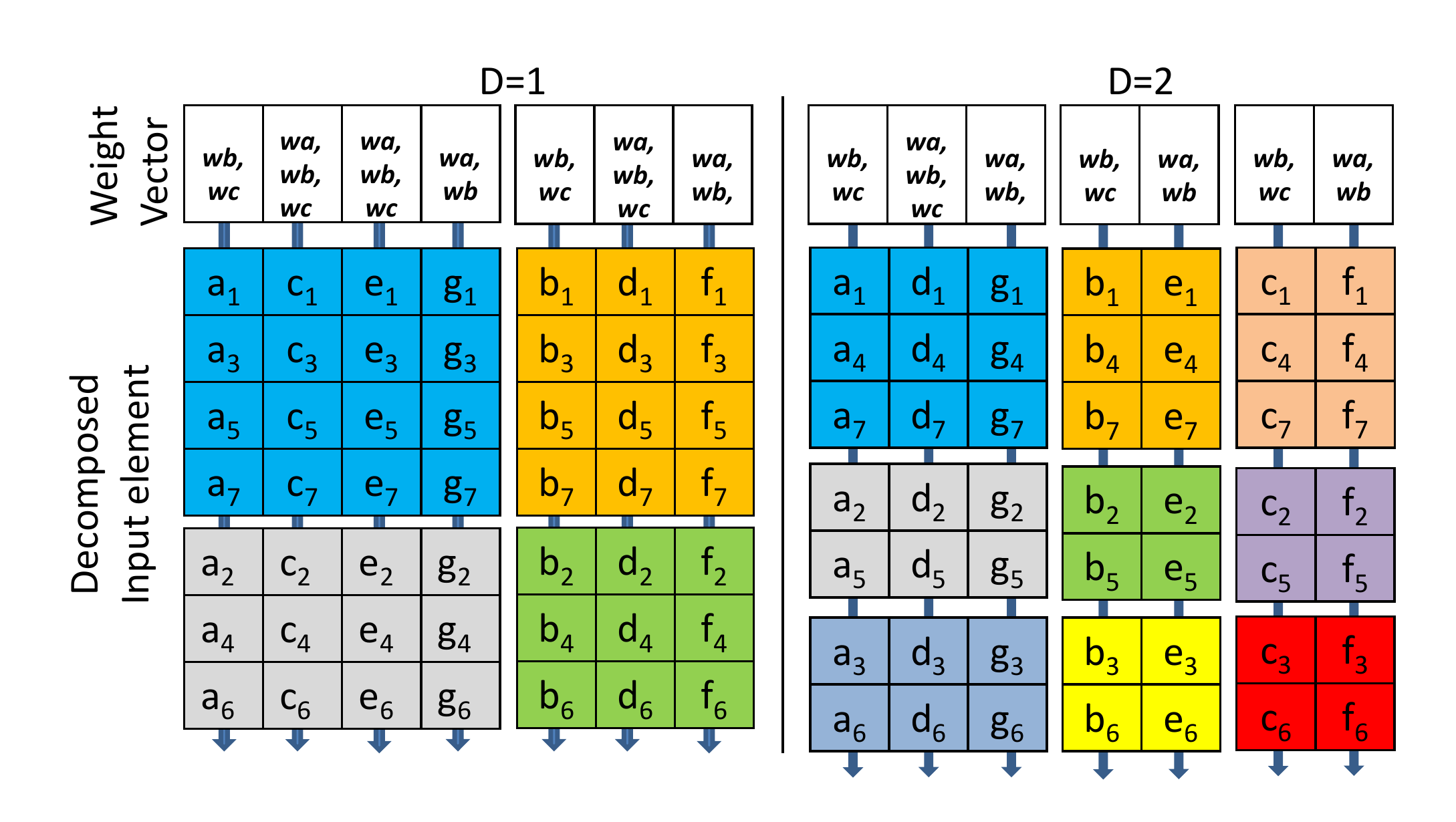}}
		\caption{Operations of each input column vector and corresponding weight column vectors in different $D$ dilated convolutions.}
		\label{fig:skip_weight}
	\end{figure}
	\begin{figure}[t]
		\centering{\includegraphics[height=60mm]{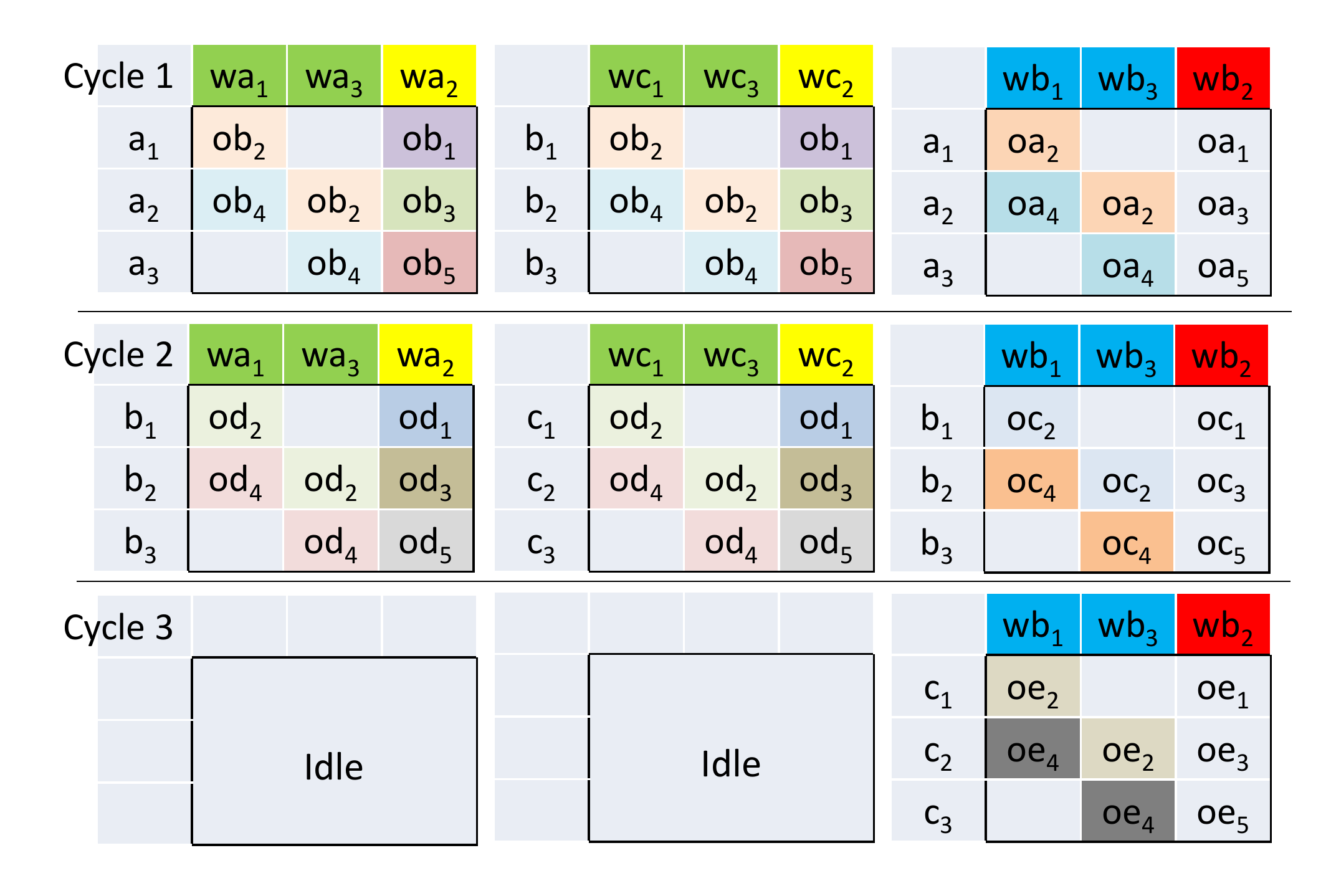}}
		\caption{Dataflow chart for the example in Fig.~\ref{fig:deconv} with 3 blocks$\times$3$\times$3 MACs. In which, the same color elements belong to the same output.}
		\label{fig:dataflow_deconv}
	\end{figure}
	
	\begin{figure}[t]
		\centering{\includegraphics[height=70mm]{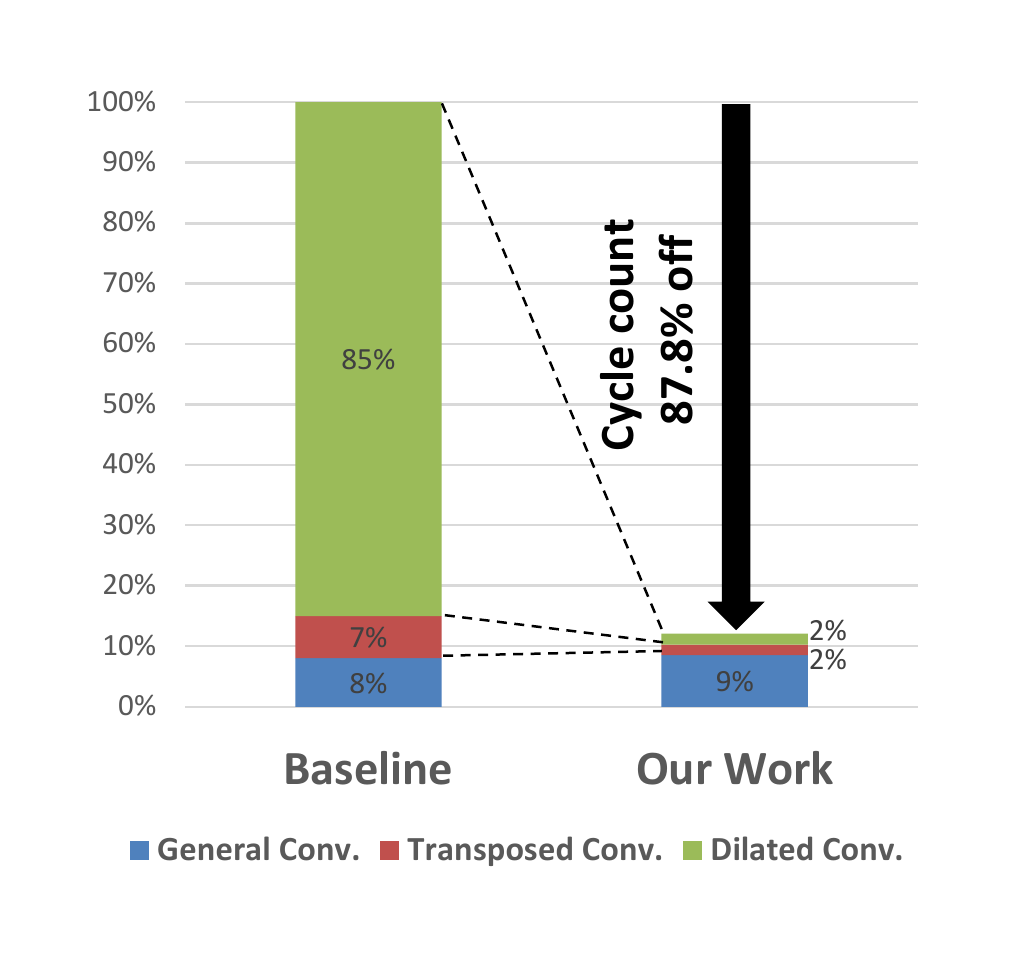}}
		\caption{The performance enhancement for our work on ENet\cite{enet}. The baseline is cycle counts on the ideal dense case. The number of MACs are the same in our work and the ideal dense case. }
		\label{fig:speedup}
	\end{figure}
	\begin{figure}[t]
		\centering{\includegraphics[height=50mm]{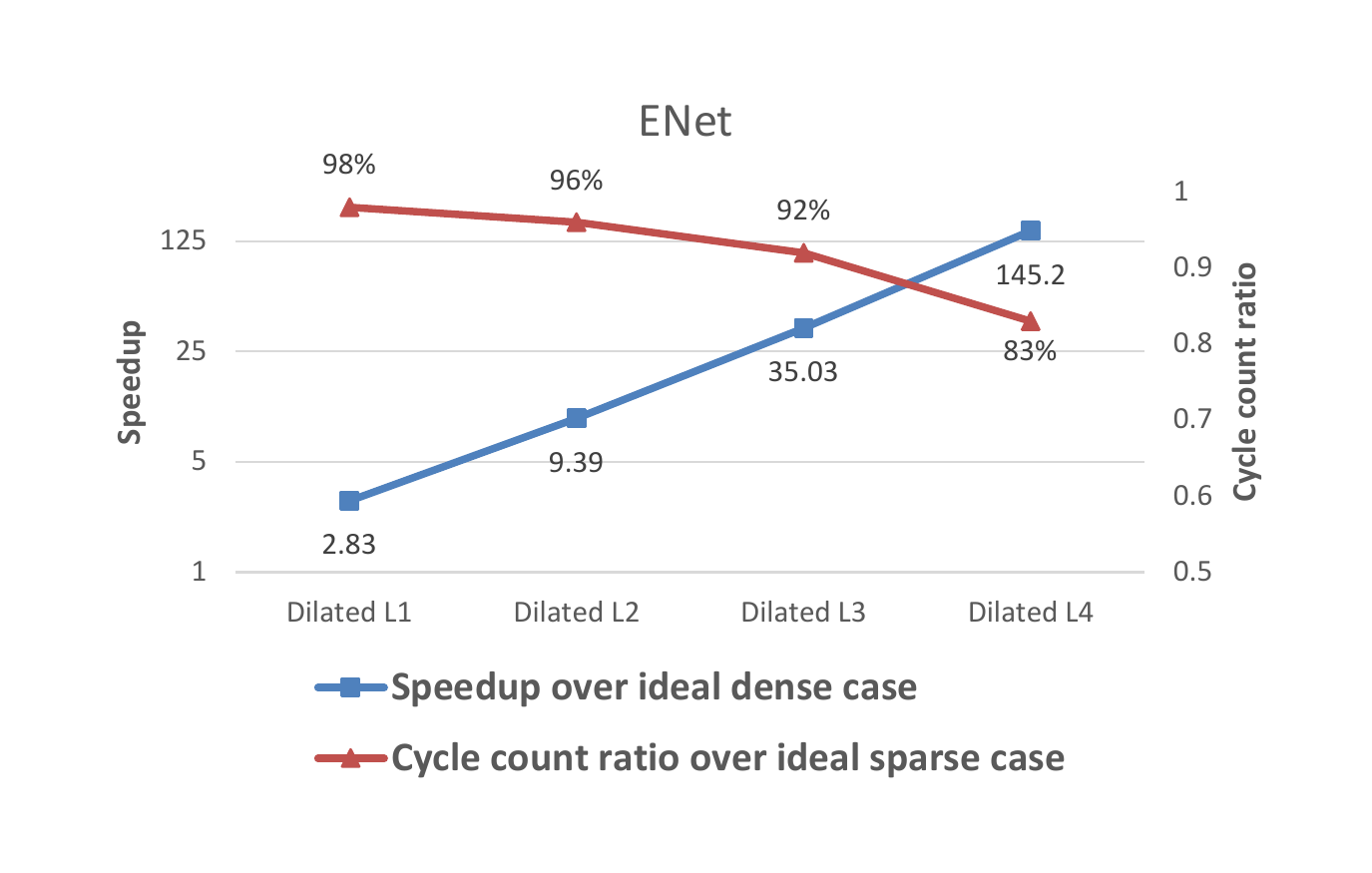}}
		\caption{The performance of dilated convolutional layers on ENet\cite{enet}. Dilated L1, L2, L3, and L4 represent dilated rate $D$ are 1, 3, 7, and 15, respectively.}
		\label{fig:enet_dilated}
	\end{figure}
	\begin{figure}[t]
		\centering{\includegraphics[height=50mm]{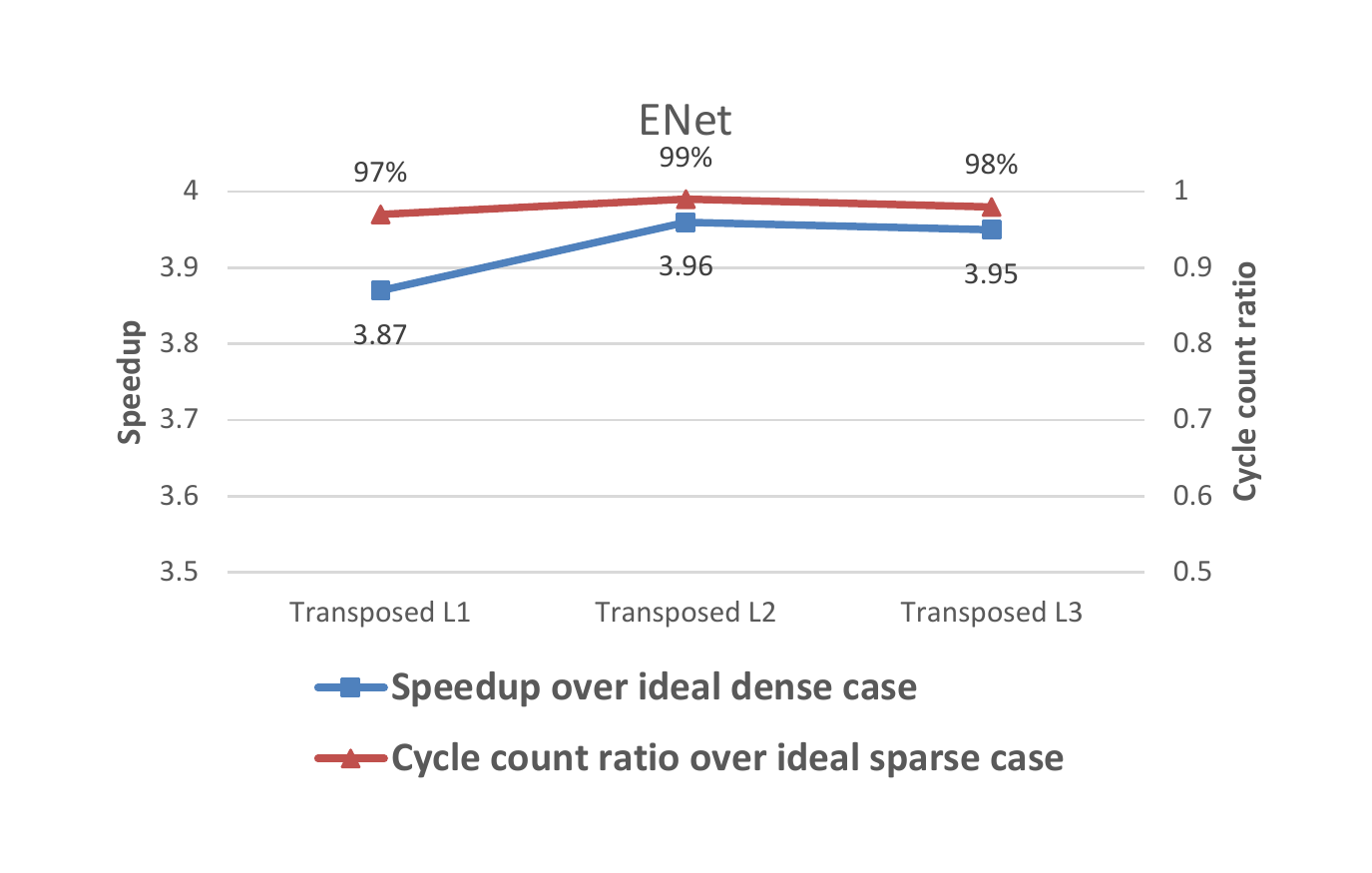}}
		\caption{The performance of transposed convolutional layers on ENet\cite{enet}. Transposed L1, L2, and L3 represent different input size are 128, 256, and 512, respectively}
		\label{fig:enet_transposed}
	\end{figure}

	\subsection{Input Decomposition for Dilated Convolutions}
	Fig.~\ref{fig:dilated_weight} shows the weight matrices with different dilation rate $D$ (that is, the different number of inserted zeros that between the adjacent weight elements). To avoid these zero computations, we decompose input of dilated convolutions into $(1+D)^2$ blocks with inspirations by \cite{deeplab}. Each block consists of input elements subsampled by $D$ from original input. Thus, as shown in Fig.~\ref{fig:decom_dilated}, the 7$\times$7 input is decomposed to 4 blocks (4$\times$4, 4$\times$3, 3$\times$4, and 3$\times$3) for $D$ = 1, or 9 blocks (3$\times$3, 3$\times$2, 3$\times$2, 2$\times$3, 2$\times$2, 2$\times$2, 2$\times$3, 2$\times$2, and 2$\times$2) for $D$ = 2. The decomposed input blocks are then followed by standard convolution (e.g. 3$\times$3 non-zero weights) to generate independent outputs. This decomposition makes dilated convolutions with any $D$ into dense convolutions and thus is suitable for all dense CNN accelerators.

	\subsection{Weight Decomposition for Transposed Convolutions}
	Fig.~\ref{fig:deconv} shows a transposed convolution example that convolves an enlarged 7$\times$7 input and 3$\times$3 weight to generate enlarged output. This convolution consists of a lot of zero computations, with only four exceptional cases as shown in Fig.~\ref{fig:deconv}. Based on this observations, we can decompose weight to four cases as shown in Fig.~\ref{fig:deconv_decom}. These four cases are four corners (2$\times$2), two horizontal endpoints (1$\times$2), two vertical points (2$\times$1), and center (1$\times$1). We decompose weight matrices to these four blocks for transposed convolutions to avoid unnecessary zero computations. Thus, the decomposed weight matrices just needs to multiply with normal input directly without zero insertion.

	\subsection{Architecture}
	The proposed method decomposes dilated and transposed convolutions into several dense normal CNN, which can be executed on a dense CNN hardware. For evaluation purpose, we apply this method to our previous proposed dense CNN architecture as shown in Fig.~\ref{fig:PE} \cite{VWA} that illustrates a $n\times3$ MAC array design of one PE block. This design is a typical systolic array type. The PE block has $n$ inputs in the same input column vector broadcasted horizontally, three weights in the same weight column vector broadcasted vertically to optimized for 3$\times$3 convolutions, and partial multiplication summed along diagonal direction. Finally, the partial sums from PE block will be accumulated to generate output in accumulator.
	
	Fig.~\ref{fig:skip_weight} shows operations of dilated convolutions on this architecture. The input column vectors in each input block will be broadcasted horizontally and the corresponding weight column vectors will be broadcasted vertically and sequentially. For boundary case, to avoid unnecessary computations due to the zero paddings at the boundary, only two weight column vectors (e.g. $\boldsymbol{wb}$, $\boldsymbol{wc}$ or $\boldsymbol{wa}$, $\boldsymbol{wb}$) are multiplied with input boundary vectors (e.g. $\boldsymbol{a}$, $\boldsymbol{g}$, $\boldsymbol{b}$, and $\boldsymbol{f}$ in $D$ = 1 and $\boldsymbol{a}$, $\boldsymbol{g}$, $\boldsymbol{b}$, $\boldsymbol{e}$, $\boldsymbol{c}$, and $\boldsymbol{f}$ in $D$ = 2). For other non-boundary input, three weight vectors are multiplied with these input vectors (e.g. $\boldsymbol{c}$, $\boldsymbol{e}$, and $\boldsymbol{d}$ in $D$ = 1 and $\boldsymbol{d}$ in $D$ = 2). The final output will be stitched together by writing the output to the target address.
	
	Fig.~\ref{fig:dataflow_deconv} shows the data flow of transposed convolutions to compute an example as in Fig.~\ref{fig:deconv}. For clarity of explanation, we assume 3 blocks with 3$\times$3 PEs in each block for Fig.~\ref{fig:dataflow_deconv}. 
	Thus, the problem is how to schedule the decomposed weight as in Fig.~\ref{fig:deconv_decom} to the architecture as in Fig.~\ref{fig:PE}. Since we have nine input ports for weight (3 blocks with 3$\times$3 PEs), one way is to assign all these nine weights in Fig.~\ref{fig:deconv_decom} to these nine input ports. In this assignment, we will assign weights that needs the same input since each 3$\times$3 PE block shares the same input. Thus, the weight assignments will be as shown in Fig.~\ref{fig:dataflow_deconv}. For this 3$\times$3 input case, it will need three cycles to complete the convolution. In which, the idle blocks at these three cycles are due to the boundary case for this small input. 
	
	\section{Experimental Result} 
	The proposed architecture has been implemented with the TSMC 40nm CMOS technology at 500MHz and simulated with the ENet\cite{enet} model trained on the Cityscapes dataset \cite{cityscapes} that is resized to 512$\times$512 as our test case.

	\subsection{Speedup}
	Fig.~\ref{fig:speedup} shows the performance enhancement with the proposed method on ENet. Our work can cut down 87.8\% of the operations by skipping zero computation. The overall speedup over the ideal dense case can reach up to 8.2X. {\bf The ideal dense case} computes all convolutions (disregarding zero or not) without considering underlying architecture constraints, which is equivalent to all multiplications and accumulations needed in the convolution.
	In above speedup, the cycle count of dilated convolutions has been reduced from 85\% to only 2\% (about 42.5X speedup) due to abundant zero computation saving. A detailed analysis shown in Fig.~\ref{fig:enet_dilated} displayed the trend of higher speedup for larger dilated rate. Fig.~\ref{fig:enet_dilated} also shows the comparison to {\bf the ideal sparse case} (only compute the nonzero elements). The presented approach has reached over 83\% to 98\% efficiency compared to the ideal sparse case. The efficiency loss is due to the zero paddings, which has more padded zeros for larger $D$ at the top and bottom of input. The cycle count of transposed convolutions has been reduced from 7\% to only 2\% (3.5X speedup).  A detailed analysis shown in Fig.~\ref{fig:enet_transposed} displays our results very close to the ideal sparse case (up to 99\%). The marginal loss is due to the tiled input. However, the cycle counts (9\%) of general convolutions in our work is a little higher than the ideal dense case (8\%) because utilization of our work is not full in the general convolutions.
	 
	\subsection{Implementation and Comparison to Other Designs}
	\begin{table}[t]
		\centering
		\caption{Implementation result and comparisons with other designs.}
		\label{table:overview}
		\begin{tabular}{|l||c|c|c|}
			\hline
			& Our work & \cite{DTCNN} & \cite{USCA} \\\hline
			Technology & 40nm & 65nm & 28nm \\\hline
			Measurements & Post-layout & Post-layout & Synthesis \\\hline
			Precision & 16 fixed & 8 & - \\\hline
			On-chip SRAM (KB) & 191 & 220.5 & 114.7 \\\hline
			Frequency (MHz) & 500 & 200 & 1449 \\\hline
			\multirow{2}{2cm}{Throughput (GOPS)$^{a}$} & 168$^{d}$/ 1377$^{e}$ & 96$^{d}$ / 639$^{e}$ & 374 \\
			& 168$^{bd}$/ 1377$^{be}$& 156$^{bd}$/ 1039$^{be}$& 261$^{bd}$\\\hline
			Supply Voltage (V) & 0.99 & 1.2 & - \\\hline
			Core Area (mm$^{2}$) & 1.5625 & 6.8  & - \\\hline
			Core Power (mW) & 155 & 196 & 201.1 \\\hline
			\multirow{2}{2.5cm}{Area efficiency(GOPS/mm$^{2}$)} & 107$^{d}$ / 881$^{e}$ & 14$^{d}$ / 94$^{e}$ & - \\
			& \bf{107$^{bd}$ / 881$^{be}$} & 23$^{bd}$ / 152$^{be}$ & - \\\hline
			\multirow{2}{2cm}{Power efficiency (TOPS/W)} & 1.08$^{d}$ / 8.88$^{e}$ & 0.49$^{d}$ / 3.26$^{e}$ & 1.86$^{d}$ \\
			& 1.08$^{cd}$ / {\bf 8.88}$^{ce}$ & {\bf 1.16}$^{cd}$/ 7.79$^{ce}$ & - \\\hline
			\multicolumn{4}	{|l|}{$^{a}$1 GMACS= 2 GOPS} \\
			\multicolumn{4}	{|l|}{$^{b}$Technology scaling ($\dfrac{process}{40nm}$)} \\
			\multicolumn{4}	{|l|}{$^{c}$Normalized~power~efficiency $=$ power~efficiency$\times(\dfrac{process}{40nm})\times(\dfrac{Voltage}{0.99V})^2$.} \\
			\multicolumn{4} {|l|}{$^{d}$The peak throughput for computing all the operations including zeros.} \\
			\multicolumn{4} {|l|}{$^{e}$The logical throughput with zero skipping on ENet \cite{enet}\cite{DTCNN}.} \\
			
			\hline                                    
		\end{tabular}
	\end{table}		
	
	Table.~\ref{table:overview} shows the implementation result and comparison with other designs dedicated to segmentation \cite{DTCNN,USCA}. The peak throughput is 168 GOPS for computing all the operations including zeros. With zero skipping, the throughput for ENet is 1377 GOPS. Our work also has much lower area cost than other designs dedicated for segmentation due to the simpler PE structure and controller. The area efficiency is 881 GOPS/mm$^{2}$ for  segmentation, which is up to 5.79X higher than \cite{DTCNN}. The power efficiency can  reach up to 8.88 TOPS/W for  segmentation which is 1.13X and 4.77X higher than \cite{DTCNN} and \cite{USCA}, respectively. The power efficiency of \cite{DTCNN} for dense CNN computation is higher than our work because of its lower bitwidth hardware to attain lower power consumption.
	
	\section{Conclusion}
	This paper proposes hardware efficient execution for dilated and transposed convolutions that decompose input or weight matrices to convert these sparse computations into dense computations. This dense computation form can be executed on a general dense CNN without extra controller overhead. Our work can cut down 87.8\% of the cycle count and 8.2X speedup over the ideal dense case. The area efficiency is up to 5.79X higher and the power efficiency is up to 4.77X than other designs for segmentation.
	
	\section*{Acknowledgment}
	This work was supported in part by the Ministry of Science and Technology, Taiwan, under Grant 109-2634-F-009-022. The authors would like to thank TSRI for its support with EDA design tools.

\end{document}